\documentclass[preprint,12pt, a4paper]{elsarticle}

\usepackage{amssymb}
\usepackage{hyperref}
\usepackage{amsmath}
\journal{SoftwareX}

\begin{document}
\renewcommand{\labelenumii}{\arabic{enumi}.\arabic{enumii}}

\begin{frontmatter}

\title{PyTomography: A Python Library for Medical Image Reconstruction}


\author[a1,a2]{L Polson}
\author[a1,a2]{R Fedrigo}
\author[a1,a2]{C Li}
\author[a1,a2]{M Sabouri}
\author[a2,a3]{O Dzikunu}
\author[a1,a2]{S Ahamed}
\author[a4]{N Karakatsanis}
\author[a6,a7]{S Kurkowska}
\author[a8]{P Sheikhzadeh}
\author[a6]{P Esquinas}
\author[a1,a2,a3,a5]{A Rahmim}
\author[a2,a5,a6]{C Uribe}

\address[a1]{
    Department of Physics \& Astronomy, University of British Columbia, Vancouver Canada
}
\address[a2]{
    Department of Integrative Oncology, BC Cancer Research Institute, Vancouver Canada
}
\address[a3]{
    School of Biomedical Engineering, University of British Columbia, Vancouver, Canada
}
\address[a4]{
   Department of Radiology, Weill Cornell Medical College, New York, U.S.A.
}
\address[a5]{
    Department of Radiology, University of British Columbia, Vancouver, Canada
}
\address[a6]{
    Molecular Imaging and Therapy Department, BC Cancer, Vancouver, Canada
}

\address[a7]{
    Department of Nuclear Medicine, Pomeranian Medical University, Szczecin, Poland
}

\address[a8]{
    Nuclear Medicine Department, IKHC, Faculty of Medicine, Tehran University of Medical Science, Tehran, Iran
}

\begin{abstract}

There is a need for open-source libraries in emission tomography that (i) use modern and popular backend code to encourage community contributions and (ii) offer support for the multitude of reconstruction techniques available in recent literature, such as those that employ artificial intelligence. The purpose of this research was to create and evaluate a GPU-accelerated, open-source, and user-friendly image reconstruction library, designed to serve as a central platform for the development, validation, and deployment of various tomographic reconstruction algorithms. PyTomography was developed using Python and inherits the GPU-accelerated functionality of PyTorch and parallelproj for fast computations. Its flexible and modular design decouples system matrices, likelihoods, and reconstruction algorithms, simplifying the process of integrating new imaging modalities using various python tools. Example use cases demonstrate the software capabilities in parallel hole SPECT and listmode PET imaging. Overall, we have developed and publicly share PyTomography, a highly optimized and user-friendly software for medical image reconstruction, with a class hierarchy that fosters the development of novel imaging applications.
\end{abstract}


\end{frontmatter}


\section*{Metadata}

\begin{table}[!h]
\begin{tabular}{|l|p{6.5cm}|p{6.5cm}|}
\hline
\textbf{Nr.} & \textbf{Code metadata description} & \textbf{Please fill in this column} \\
\hline
C1 & Current code version & v3.2.3 \\
\hline
C2 & Permanent link to code/repository used for this code version & \url{https://github.com/qurit/PyTomography} \\
\hline
C3  & Permanent link to Reproducible Capsule &  None. Reproducible examples provided as part of the documentation.\\
\hline
C4 & Legal Code License   & MIT License \\
\hline
C5 & Code versioning system used & git \\
\hline
C6 & Software code languages, tools, and services used & python \\
\hline
C7 & Compilation requirements, operating environments \& dependencies & python$\geq
$3.9, numpy$\geq$1.24.2, scipy$\geq$1.10.1, pydicom$\geq$2.0.0, kornia$\geq$0.6.12, torch$\geq$1.10.2, fft-conv-pytorch$\geq$1.2.0, matplotlib$\geq$3.8.0, uproot$\geq$5.2.1, rt\_utils$\geq$1.2.0, nibabel$\geq$5.1.0, h5py$\geq$3.11.0, torchrbf, pandas$\geq$2.2 \\
\hline
C8 & If available Link to developer documentation/manual & \url{https://pytomography.readthedocs.io/en/latest/} \\
\hline
C9 & Support email for questions & \url{https://pytomography.discourse.group/}\\
\hline
\end{tabular}
\caption{Code metadata (mandatory)}
\label{codeMetadata} 
\end{table}

\section{Motivation and significance}

Medical imaging forms a cornerstone in modern healthcare by providing visual and quantitative information about internal body structures and functions. It enables early detection \cite{early_detection}, accurate diagnosis, and precise treatment planning for a wide range of medical conditions \cite{treatment_planning,treatmentplanning2,treatmentplanning3}, contributing to improved patient outcomes and enhanced medical decision-making. Reconstruction algorithms are routinely used to generate 3D images that can be used for both research and clinical decision making \cite{iterative_paper}.\\

While the development and validation of reconstruction techniques remains an active field of research \cite{hermes_spect_ap, attenuationless_CT, deep_image_prior, spect_proj_interp}, it can often be difficult to share and disseminate one's findings. Additionally, while manufacturers of tomographic imaging equipment such as Single Photon Emission Computed Tomography (SPECT) and Positron Emission Tomography (PET) cameras provide their own internal reconstruction software, users cannot access all implementation details; this limits reproducibility in scientific studies. These issues could be solved by moving to an open source paradigm for medical image reconstruction. Here, the imaging community could openly collaborate in the development of reconstruction techniques. Adoption of a singular open-source framework, as opposed to private frameworks developed independently by various research groups, would strengthen the ability to develop, share, and compare findings. This could ultimately help accelerate the development and translation of new diagnostic imaging capabilities into clinical practice.\\

A few tools have already been developed to try and address some of these issues. STIR \cite{STIR} and CASToR \cite{castor} are two C/C++ based open source and collaborative image reconstruction frameworks that support a variety of reconstruction algorithms for SPECT and PET. NiftyPET \cite{markiewicz2018niftypet} is a python library which employs GPU-accelerated code written in CUDA C for PET reconstruction. QSPECT \cite{QSPECT} and TIRIUS support use of the ordered subset expectation maximum (OSEM) algorithm for SPECT and PET reconstruction respectively. Other libraries, such as TomoPy \cite{tomopy} and the ASTRA toolbox \cite{astra}, support a variety of GPU-accelerated image tomographic reconstruction algorithms, although they lack extensive SPECT/PET system modeling. Web links to all these tools, as well as a list of other available tools for image reconstruction, can be found on the NMMItools webpage of the Society of Nuclear Medicine and Molecular Imaging (SNMMI).\\

The unique purpose of our present research and effort is to build upon recent efforts in GPU-accelerated reconstruction \cite{gpu_paper_2010, belzunce2012cuda, ha_2013, schramm2022parallelproj} and AI-based reconstruction \cite{attenuationless_CT, deep_image_prior, spect_proj_interp} to establish an open source medical image reconstruction programming framework written in a popular programming language that facilitates collaboration. Valuable past efforts have focused on CUDA programming in languages such as C/C++ to accelerate reconstruction times in SPECT \cite{gpu_paper_2010} and PET \cite{gpu_paper_2010, belzunce2012cuda} reconstruction through the use of multi-threading. While some contributors to an open-source framework may be comfortable with C/C++ programming, others may be unfamiliar with these languages; any tool that prioritizes community contributions should be designed with a front end application programming interface (API) that permits either of these options. Some of the advantages of a python front end API are that (i) it is a widely known programming language and thus facilitates community use, (ii) it enables direct interaction with other python libraries such as PyTorch for deep learning, and (iii) it enables backend development and contributions from other programming languages if necessary. This may address limitations of the C/C++ front end libraries STIR and CASToR, which are currently the most popular open-source reconstruction frameworks in emission tomography.\\

To this end, we developed the python library PyTomography with the priorities to: (i) implement standard and traditional imaging modalities and reconstruction algorithms, (ii) disseminate recent research developments, such as the deep image prior \cite{deep_image_prior}, and (iii) encourage  community involvement via extensive documentation and user-friendly tutorials. Based on the motivations previously described, the main functionality of the library has been developed using PyTorch. Other libraries are used for more particular tasks; the python library paralleproj \cite{schramm2022parallelproj} is used for line integral computation in PET projectors. While the present focus of PyTomography is the development and validation of reconstruction algorithms for SPECT and PET, implementation of other imaging modalities and reconstruction algorithms can be added using the building blocks provided. \\

In addition to the examples provided in this paper, PyTomography has support for high energy SPECT reconstruction via interface with the SPECTPSFToolbox \cite{polson2024fastaccuratecollimatordetectorresponse}, reconstruction using multiple SPECT photopeaks simultaneously, and region-based uncertainty estimation \cite{polson2024uncertaintypropagationprojectionsregion}; such features are not available in the other open-source libraries. There is ongoing work in the development of an associated graphical user interface extension in 3D slicer for SPECT reconstruction \cite{3DSlice}; the corresponding repository can be found on the PyTomography Github organization page.  \\

Overall, the aim of this study is to introduce the community to the PyTomography software project, demonstrate its current capabilities and flexibility, and to encourage community involvement in the continuation of its development. Example applications are explored in SPECT and PET reconstruction, though the library also has capabilities for clinical CT reconstruction as well. In what follows, we elaborate on our methods and results for PyTomography.

\section{Software description}

\subsection{Software scope and functionality}
A brief overview of the mathematical paradigm of medical imaging is presented in order to motivate the software architecture. The mathematical conventions used in this paper are as follows. Vectors are represented by lower case letters, while linear operators are represented by upper case letters. The product of two vectors $vw$ and division of two vectors $v/w$ are inferred to be point-wise operations. Components of a vector are given by $v_i$, and components of linear operators are given by $A_{ij}$. When $A1$ or $A^T1$ is written, ``1'' is to be interpreted as a vector (in the appropriate vector space) with all elements equal to 1. When classes/functions are quoted, they refer to the either (i) functionality of the current PyTomography version or (ii) functionality of dependencies (see \verb|pyproject.toml| file) of the current PyTomography version.

\subsubsection{Projectors}\label{sec:trans_proj}

In the paradigm of tomographic medical imaging, there are two main vector spaces: the ``projection'' space $\mathcal{V}$ consisting of measured data $g$ and the ``object'' space $\mathcal{U}$ consisting of 3D objects $f$. Since measured and reconstructed data is digitized, $\mathcal{U}$ and $\mathcal{V}$ are assumed here to be finite dimensional vector spaces. In SPECT imaging, for example, $\mathcal{V}$ consists of a set of 2D pixelated images acquired by counting photons emitted by radiopharmaceuticals, and $\mathcal{U}$ represents the voxelized spatial distribution of radioactivity concentration in the object being imaged.\\

Linear imaging systems can be characterized by a system matrix $H:\mathcal{U} \to \mathcal{V}$. Its components $H_{ij}$ represent the contribution of element $j$ in object space to element $i$ in projection space. In emission tomography, analytical modeling of the data expectation $\bar{g}$ can be achieved via 
$\bar{g} = Hf + s$ where $s$ is an additive term containing estimates for scattered events (SPECT) or scatter and random coincidences (PET). The operation of $H$ on an object vector is known as forward projection (FP), while the operation of $H^T$ on a projection vector is known as back projection (BP). Example implementations available in PyTomography are listed below:

\begin{itemize}
    \item \textbf{Parallel Collimator SPECT}: Modeling of primary detections in this imaging modality can be approximated using the ``rotate+sum'' technique \cite{gpu_paper_2010}; in this case, the system matrix can be written as
    \begin{equation}
        H = \sum_n v_n \otimes \left( \mathcal{T}_{\text{fov}}^{(n)}~ \mathcal{P}_x~ \mathcal{T}_{\text{cdr}}^{(n)} \mathcal{T}_{\text{att}}^{(n)}~ \mathcal{T}_{\text{trans}}^{(n)} ~ \mathcal{T}_{\text{rot}}^{(n)} \right)
        \label{eq:SPECT_SM}
    \end{equation}
    where $n$ designates the projection angle, $\mathcal{T}_{\text{rot}}^{(n)}$ implements a rotation, $\mathcal{T}_{\text{trans}}^{(n)}$ implements a translation that accounts for non-central projections, $\mathcal{T}_{\text{att}}^{(n)}$ implements the adjustment for photon attenuation, $\mathcal{T}_{\text{cdr}}^{(n)}$ implements depth-dependent collimator blurring, $\mathcal{P}_x$ is summation along the $+x$ axis, $ \mathcal{T}_{\text{fov}}^{(n)}$  applies a mask in projection space that accounts for a finite field of view, $v_{n}$ is a unit vector of length $N$ with 1 in component $n$, and $\otimes$ is the outer product. The most computationally expensive operations in this sequence are $\mathcal{T}_{\text{cdr}}^{(n)}$ and $\mathcal{T}_{\text{rot}}^{(n)}$, which are implemented using the PyTorch functions \verb|rotate| (from \verb|torchvision.transforms.functional|) and \verb|conv1d|/\verb|conv2d| (from \verb|torch.nn|).
     
    \item \textbf{PET}: Modeling of primary events (no randoms or scatters) in PET can be approximated as
    \begin{equation}
        H = \mathcal{T}_{\eta} ~ \mathcal{T}_{\omega} ~ \mathcal{P}_{\Sigma} ~ \mathcal{T}_{rr} 
        \label{eq:PET_SM}
    \end{equation}
    where $\mathcal{T}_{rr}$ implements object-space resolution modeling (typically via a uniform Gaussian kernel), $\mathcal{P}_{\Sigma}$ implements line integral computation (listmode/sinogram), $\mathcal{T}_{\eta}$  adjusts for detector normalization, and $\mathcal{T}_{\omega}$ adjusts for attenuation effects. The most computationally expensive operation is $\mathcal{P}_{\Sigma}$, which is implemented using \verb|joseph3d_fwd| (non-TOF) and \verb|joseph3d_fwd_tof_sino|/ \verb|joseph3d_fwd_tof_lm| (TOF) from parallelproj.
\end{itemize}

PyTomography can be used to construct projectors that extend projection space using the \verb|ExtendedSystemMatrix| class; such functionality is necessary, for example, when modeling multiple photopeaks in SPECT or when using motion correction. In such situations, the extended projection space has dimensionality $(N,...)$ where $...$ is the original dimensionality of projection space, and $N$ represents the number of photopeaks or time gates during acquisition. Mathematically, these projectors can be written as
\begin{equation}
    H' = \sum_n v_n \otimes B_n H_n A_n
    \label{eq:extended_sm}
\end{equation}
where $\left\{A_n\right\}$ and $\left\{B_n\right\}$ are additional operations, $\left\{H_n\right\}$ are a sequence of system matrices, $v_n$ is a basis vector of length $N$ with a value of $1$ in component $n$, and $\otimes$ is an outer product. This can be utilized in the following examples:

\begin{itemize}
    \item \textbf{Periodic Motion Correction}: $A_n$ represent spatial deformation operators corresponding to time gate $n$, $H_n$ represent system matrices for each motion transform (attenuation modeling may depend on object deformation), and $B_n=1$. This maps to an extended projection space $(N,...)$ where $N$ is the number of different time gates (e.g. during a breathing cycle).
    \item \textbf{Multiple SPECT Photopeaks}: In this scenario, $A_n$ are scaling factors that adjust for detector sensitivities and branching ratios between different emission peaks, $B_n$ are identity mappings, and $H_n$ represents the system matrix corresponding to a particular photopeak. The extended projection space has dimensions $(N,...)$ where $N$ is the number of different photopeaks.
\end{itemize}

\subsubsection{Likelihoods}

A statistical likelihood function $\tilde{L}(g|\bar{g})$ is proportional to the probability of observing an actual realization of data $g$ given expectation $\bar{g}$. Since $\bar{g}$ is approximated using $f$ and $H$, it follows that $\tilde{L}(g|\bar{g}) = \tilde{L}(g|f,H,...)$, where the ellipsis ``$...$'' contains additional parameters such as a scatter estimate $s$; the ellipsis will be assumed implicit for the remainder of this paper. Many iterative reconstruction algorithms seek to maximize the likelihood function via gradient ascent by exploiting the gradient $\nabla_f \tilde{L}(g|f,H)$. Since (i) computation of $\nabla_f \log \tilde{L}(g|f,H)$ is more tenable than $\nabla_f \tilde{L}(g|f,H)$ and (ii) maximization of $\log \tilde{L}(g|f,H)$ is equivalent to maximization of $\tilde{L}(g|f,H)$, the statistical ``log-likelihood'' is nearly always considered instead. In PyTomography, the ``likelihood'' $L(g|f,H)$ refers to an objective function to be maximized; as such, the ``likelihood'' $L(g|f,H)$ is often used interchangeably with the statistical ``log-likelihood'' $\log \tilde{L}(g|f,H)$ for a particular modality. Likelihood functions for different modalities are shown in Table \ref{tab:sample_classes}; evaluation of a likelihood and its derivatives require forward and back projection of system matrices.

\subsubsection{Reconstruction Algorithms}\label{sec:recon_algos}

An image reconstruction algorithm $A$ uses measured data $g$  (i.e. projections) to estimate a corresponding object $f$ that would produce this $g$ given a system matrix $H$. Statistical algorithms find the image estimate $f$ which maximizes the likelihood function $L(g|H,f)$ characterizing the imaging system. This can be expressed as

\begin{equation}
    f = A(L,...)
    \label{eq:generic_recon}
\end{equation}
where the ellipsis ``$...$'' includes all additional hyperparameters required for the algorithm; the ellipsis will be assumed implicit for the remainder of this paper. The main algorithms in PyTomography are preconditioned gradient ascent (PGA) algorithms. These algorithms aim to maximimize the likelihood via iterative updates that
take the form

\begin{equation}
    f^{(n+1)} = f^{(n)} + C^{(n)}(f^{(n)},H^{(n)}) \left[\nabla_{f} L(g^{(n)}|f^{(n)},H^{(n)}) - \beta^{(n)} \nabla_{f} V(f^{(n)}) \right]
    \label{eq:PGA}
\end{equation}

where $n$ is the subiteration index (used as a superscript and not an exponent for all variables), $f^{(n)}$ is the object estimate of subiteration $n$, $g^{(n)}$ specifies a subset of data used in subiteration $n$, $H^{(n)}$ is a system matrix projecting only to the current subset, $C^{(n)}(f^{(n)})$ is a preconditioner factor that depends on the object estimate, $V(f^{(n)})$ is a prior function used to regularize the algorithm, and $\beta^{(n)} = c^{(n)} \beta$ is a scaling factor for the prior function (scaled by the fraction of subset data $c^{(n)}$). These algorithms partition the data $g$ into $M$ approximately equal subsets; a single iteration of an algorithm is defined as the number of subiterations where all subsets have been used once. Various PGA algorithms available in PyTomography are shown in Table \ref{tab:sample_classes}; they are characterized by their preconditioner factor.\\

Prior functions $V(f)$ used in reconstruction often take the form
\begin{equation}
    V(f) = \sum_r \sum_s w_{rs} \phi(f_r, f_s)
\end{equation}
where $w_{rs}$ provides a weighting factor between the locations voxel $r$ and voxel $s$ and $\phi(f_r, f_s)$ characterizes a weighting factor between the observed image values at voxel $r$ and voxel $s$. When $w_{rs}$ is non-zero only for neighbouring voxels, the prior is known as a nearest neighbour prior; various nearest neighbour priors implemented in PyTomography are shown in Table \ref{tab:sample_classes}. While the weighting factor $w_{rs}$ is typically proportional to inverse Euclidean distance, it may also be dependent on an additional anatomical image $\chi$ whereby $w_{rs}=w_{rs}(\chi_r, \chi_s)$  Since anatomical information is used, this is commonly referred to as an anatomical prior (AP); variations of this technique have recently become popular in clinical practice due to enhanced lesion quantitation and detectability in bone SPECT/CT \cite{hermes_spect_ap}. In this paper, use of an AP refers to using only the 8 most similar neighbours based on absolute differences of HU values in a CT scan or attenuation map. The prior used in subsequent examples in this paper is the relative difference penalty (RDP) \cite{RDP}, since it is commonly used in vendor reconstruction.\\

\begin{table}[h]
    \centering
    \caption[asd]{Sample PyTomography classes and their corresponding identifying features. $s$ corresponds to additive term in likelihoods. The sample algorithms shown are the ordered subset expectation maximum (OSEM) \cite{osem}, ordered subset maximum a posteriori one step late (OSMAPOSL) \cite{OSL}, block sequential regularized expectation maximum (BSREM) \cite{BSREM}. $n$ is used as a superscript (not an exponent) for all variables. For BSREM, $\alpha^{(n)}$ is a relaxation parameter. The sample nearest neighbour prior functions are characterized by their image weighting function $\phi(f_r, f_s)$. \label{tab:sample_classes}}
    \begin{tabular}{|c|c|}
        \hline
        \textbf{Projector classes} & System matrix $H$\\
        \hline
        \hline
        \verb|SPECTSystemMatrix| & Equation \ref{eq:SPECT_SM}\\
        \verb|PETLMSystemMatrix|  & Equation \ref{eq:PET_SM} \\
        \hline
        \hline
        \textbf{Likelihood classes} & Gradient $\nabla_f L(g|f,H)$\\
        \hline
        \hline
        \verb|PoissonLogLikelihood| & $H^T \left( \frac{g}{Hf+s} \right) - H^T 1$\\
        \verb|NegativeMSELikelihood|  & $H^T W (g - (Hf+s))$ \\
        \hline
        \hline
        \textbf{Algorithm classes} & Preconditioner factor $C^{(n)}(f^{(n)},H^{(n)})$\\
        \hline
        \hline
        \verb|OSEM| & $f^{(n)} / \left(\left(H^{(n)}\right)^T 1\right)$ and $V(f^{(n)})=0$\\
        \verb|OSMAPOSL| & $f^{(n)} / \left(\left(H^{(n)}\right)^T 1 + \nabla_f V(f^{(n)})\right)$ \\
        \verb|BSREM| & $f^{(n)} \alpha^{(n)} / \left(c^{(n)} H^T 1\right)$\\
        \hline
        \hline
        \textbf{Prior classes} & Weighting function $\phi(f_r, f_s)$\\
        \hline
        \hline
        \verb|QuadraticPrior| & $\left((f_r-f_s)/\delta \right)^2 /4 $\\
        \verb|LogCoshPrior| & $\tanh((f_r-f_s)/\delta)$ \\
        \verb|RelativeDifferencePrior| & $\left((f_r-f_s)^2\right)/\left(f_r+f_s+\gamma|f_r-f_s|\right)$\\
        \hline
    \end{tabular}
\end{table}

PyTomography also offers support for the Deep Image Prior Reconstruction (DIPRecon) algorithm. Details regarding the implementation of this algorithm can be found in Algorithm 2 of Gong et.\ al \cite{deep_image_prior}. The purpose of this reconstruction technique is to use anatomical images (CT/MR) to enhance functional images (SPECT/PET) through a sequence of neural network overfitting and mixing with OSEM updates. This reconstruction algorithm requires the use of an anatomical image $z$ and a neural network architecture $\mathcal{N}(\cdot;\theta)$ that can be trained to predict a reconstructed functional image $f$ given $z$ as input. This paper explores the use of DIPRecon on an ultra-high resolution PET/MR brain phantom. 

\subsection{Software architecture}

\begin{figure}[ht]
   \begin{center}
   \includegraphics[width=13cm]{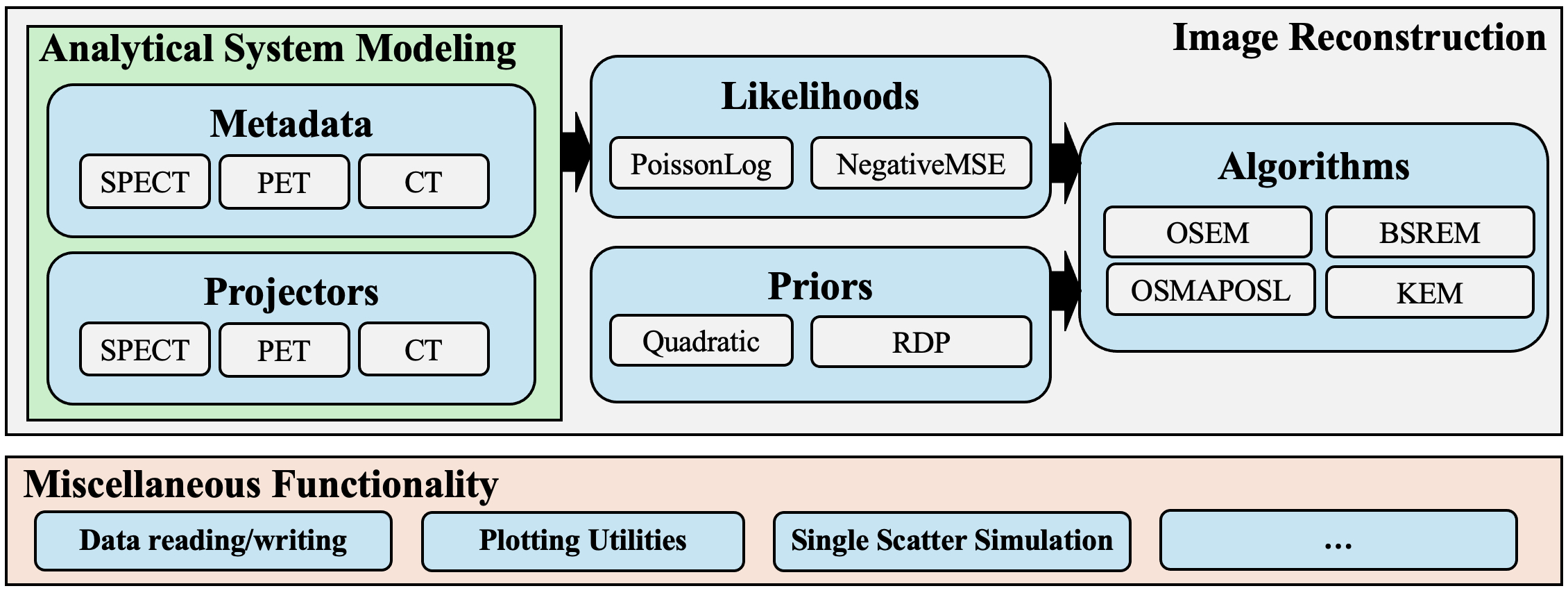}
   \caption{Software architecture diagram for PyTomography. Metadata and projectors consist of an independent component of the library that can be used for analytical system modeling, but can also be used to construct likelihoods that can be passed to image reconstruction algorithms. Implementations of various components displayed here can be found in Table \ref{tab:sample_classes}.
   \label{fig:software_diagram} 
    }  
    \end{center}
\end{figure}

PyTomography has been carefully designed to permit arbitrary implementations of system matrices (projectors) which can be used in various likelihood functions that themselves interface with reconstruction algorithms. The components of PyTomography were designed to facilitate this architecture and are listed below.  A visual representation of the software library along with sample code representing a SPECT reconstruction use case is shown in Figure \ref{fig:software_diagram}.

\begin{enumerate}
    \item \textbf{Data/Metadata}. The standard data type used to represent images and raw projection data is the tensor class of PyTorch; data can be stored on CPU or GPU. Metadata classes contain all auxiliary information needed for image reconstruction pertaining to a particular imaging system, such as voxel size, spatial dimensions, and information pertaining to the raw data of a particular modality.  The input/output (I/O) module is used for reading various file formats; it currently supports SIMIND, DICOM (SPECT/CT), GATE (PET), and HDF5 listmode (PET) formats. Data reading consists of loading the generated files and constructing the appropriate data/metadata instances within the PyTomography framework. The I/O module is also used to read internal files of the library, such as coefficients to convert Hounsfield Units to attenuation coefficients, files containing SPECT collimator information, and files containing PET scanner geometry.
    \item \textbf{Projectors}. Projectors are mathematical operators that map between object and projection space. The system matrix is an example of a projector that models a particular imaging system. While the input/output datatype to projectors is the PyTorch tensor class, internal code can utilize functionality from other libraries. Projector subclasses inherit from a generic \verb|SystemMatrix| class and implement a \verb|forward|, \verb|backward|, and \verb|compute_normalization_factor| method. Projectors used in subset based algorithms should additionally implement \verb|set_n_subsets|, \verb|get_projection_subset| and \verb|get_weighting_subset|; the second method takes in projection data and returns subset data based on how the projector defines subsets, and the third gets the corresponding weight ($c^n$) for scaling preconditioners.
    \item \textbf{Likelihoods}. Likelihoods specify the likelihood function that characterizes the acquired data. Likelihoods $L(g|H,f)$ are initialized using a system matrix $H$, projection data $g$, and any additional dependencies; once initialized they act as functions of an image estimate $f$. Likelihoods inherit from a generic \verb|Likelihood| class and should implement the \verb|compute_gradient| method. Higher order derivatives of the likelihood may also be defined as separate methods within the class.
    \item \textbf{Algorithms}. Algorithms are used for image reconstruction, and are designed to interface with all imaging modalities. Statistical iterative algorithms require as input a likelihood instance, and may take as optional arguments prior functions and callbacks. Linear PGAAs inherit from a generic \verb|LinearPreconditionedGradientAscentAlgorithm| class and simply need to implement the \verb|_linear_preconditioner_factor| and optionally re-implement the \verb|__init__| method to account for any additional required arguments.
\end{enumerate}

\begin{figure}[ht]
   \begin{center}
   \includegraphics[width=5.5cm]{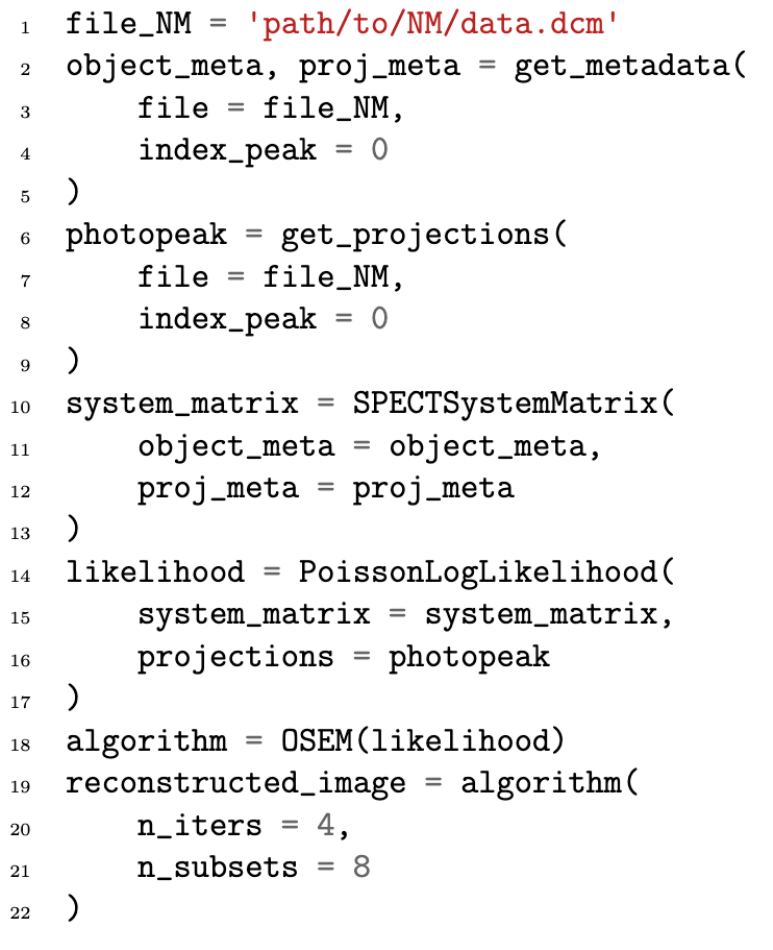}
   \caption{Sample code for a simplified SPECT reconstruction use case (e.g.\ no attenuation/PSF/scatter correction) that demonstrates the modular structure of the library.
   \label{fig:code} 
    }  
    \end{center}
\end{figure}

Figure \ref{fig:software_diagram} shows a graphical depiction of the structure of the library. Sample code demonstrating the modularity of the library is shown in Figure \ref{fig:code} for a simplified SPECT use case (e.g.\ no attenuation, PSF, or scatter correction). Lines 1-9 demonstrate reading the data/metadata, lines 10-13 demonstrate constructing the projector using the loaded data/metadata, lines 14-17 demonstrate creating the likelihood using the system matrix and loaded data, and lines 18-22 demonstrate creating a reconstruction algorithm and obtaining a reconstructed image. A new contribution to the library might involve adding a new system model and replacing lines 1-13 while reusing lines 14-22. Another possible contribution might include a new reconstruction algorithm; in this case lines 18-22 would need to be replaced while the SPECT projector code can be reused. 

\section{Illustrative examples}

Multiple examples are explored to highlight the capabilities of PyTomography. Specifically, reconstruction of parallel-collimator SPECT and LM-TOF PET data is demonstrated, but it should be emphasized that these are only intended as example use cases for particular imaging modalities. Validation against other software libraries is also shown in the Appendix, which contains (i) SIMIND simulated SPECT data reconstruction validated against STIR,  (ii) DICOM SPECT data (obtained on a GE Tandem Discovery 670 Pro) validated against MIM's SPECT reconstruction functionality, and (iii) GATE PET data reconstruction validated against CASToR. We chose to validate DICOM reconstruction against MIM, and not the GE internal reconstruction software, because it represents an external library with access to the same information as PyTomography.\\

All computations were performed using a Microsoft Azure virtual machine (Standard NC6s v3) with a 6 CPUs (Intel(R) Xeon(R) CPU E5-2690 v4 @ 2.60GHz), 112 GB of RAM, and a TeslaV100 GPU. Python scripts and notebooks used to obtain these results can be found at \url{https://github.com/qurit/PyTomography_paper_code}.\\

\subsection{SPECT}
This section highlights SPECT reconstructions using available reconstruction algorithms in PyTomography, demonstrating differences in quantitative metrics between different algorithms. Two examples are considered: (i) reconstruction of clinical SPECT data and (ii) reconstruction of Monte Carlo SPECT data generated via SIMIND. In the clinical case, the patient received radiopharmaceutical therapy with $^{177}$Lu-PSMA-617 for prostate cancer and was imaged 3 hours post injection; the XCAT phantom used as input for SIMIND was configured such that the kidney, liver, and lesion activity concentrations roughly approximated the clinical patient. In each case, the data correspond to acquisition on a GE Tandem Discovery 670 Pro SPECT system (MEGP collimators) with 60 projections (25~s per projection). The clinical data consists of acquisition at two bed positions with projections of shape $128 \times 128$, while the simulated data consists of acquisition at one bed position with projections of shape $128 \times 240$. The two lesions inserted in the XCAT phantom were (i) based on an observed bone metastasis in the sternum of the clinical patient, and (ii) placed in the liver. In each case, the system matrix employed attenuation and PSF modeling, and the likelihood function considered an additive scatter term obtained via the triple energy window technique. The reconstruction algorithms considered were OSEM (4it/6ss), BSREM-RDP (40it/6ss), BSREM-RDPAP (40it/6ss), and KEM (10it/6ss) [`it' for iterations; and `ss' for subsets]; algorithms using RDP used the hyperparameters $\beta=0.3$, $\gamma=2$, RDPAP only considered the 8 nearest neighbours based on the attenuation map, and KEM used Eq.\ 5 of Vuohijoki\ \cite{hermes_spect_ap} for its filter. The required computation times are shown in Table \ref{tab:recon_times}.\\

\begin{figure}[ht]
   \begin{center}
   \includegraphics[width=13.5cm]{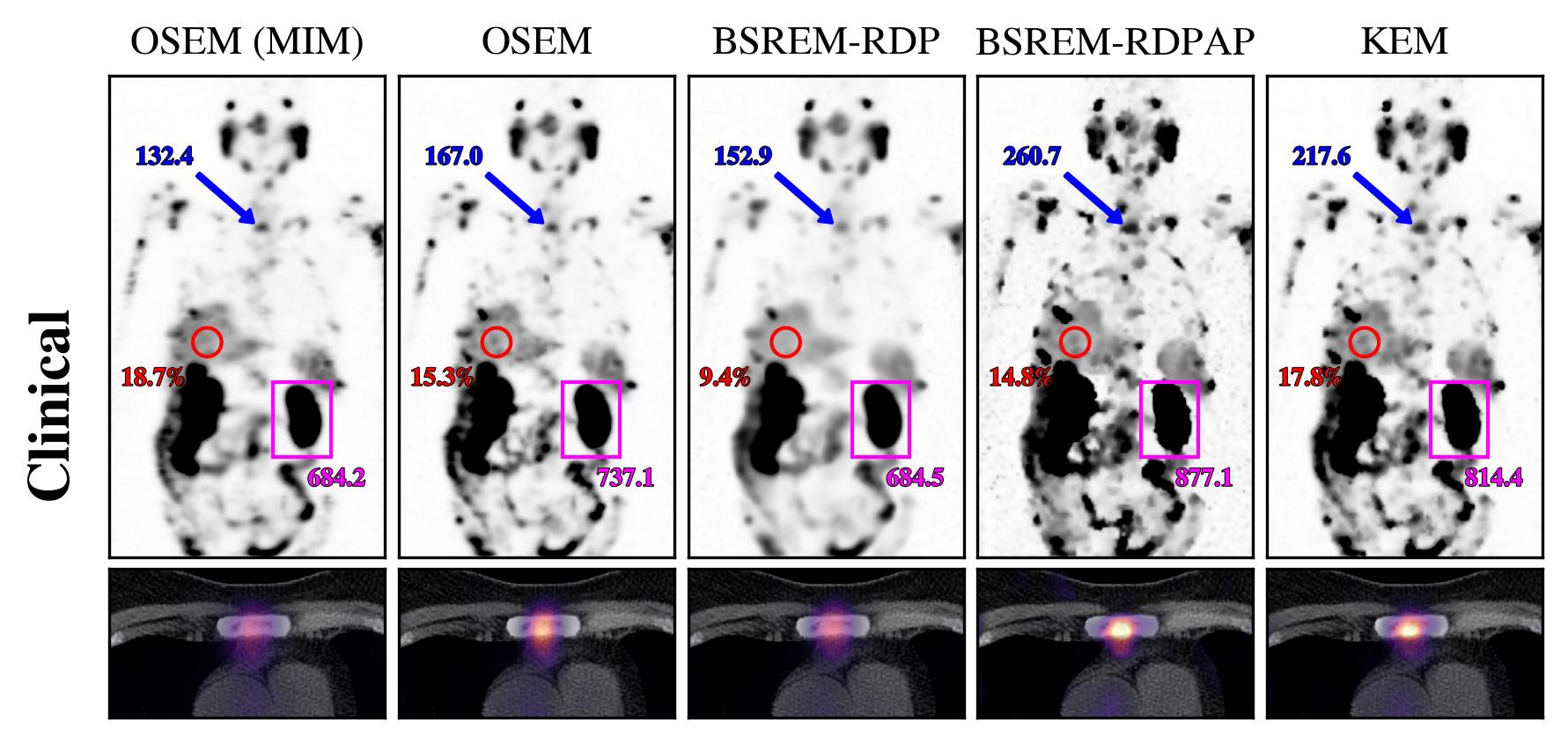}
   \includegraphics[width=13.5cm]{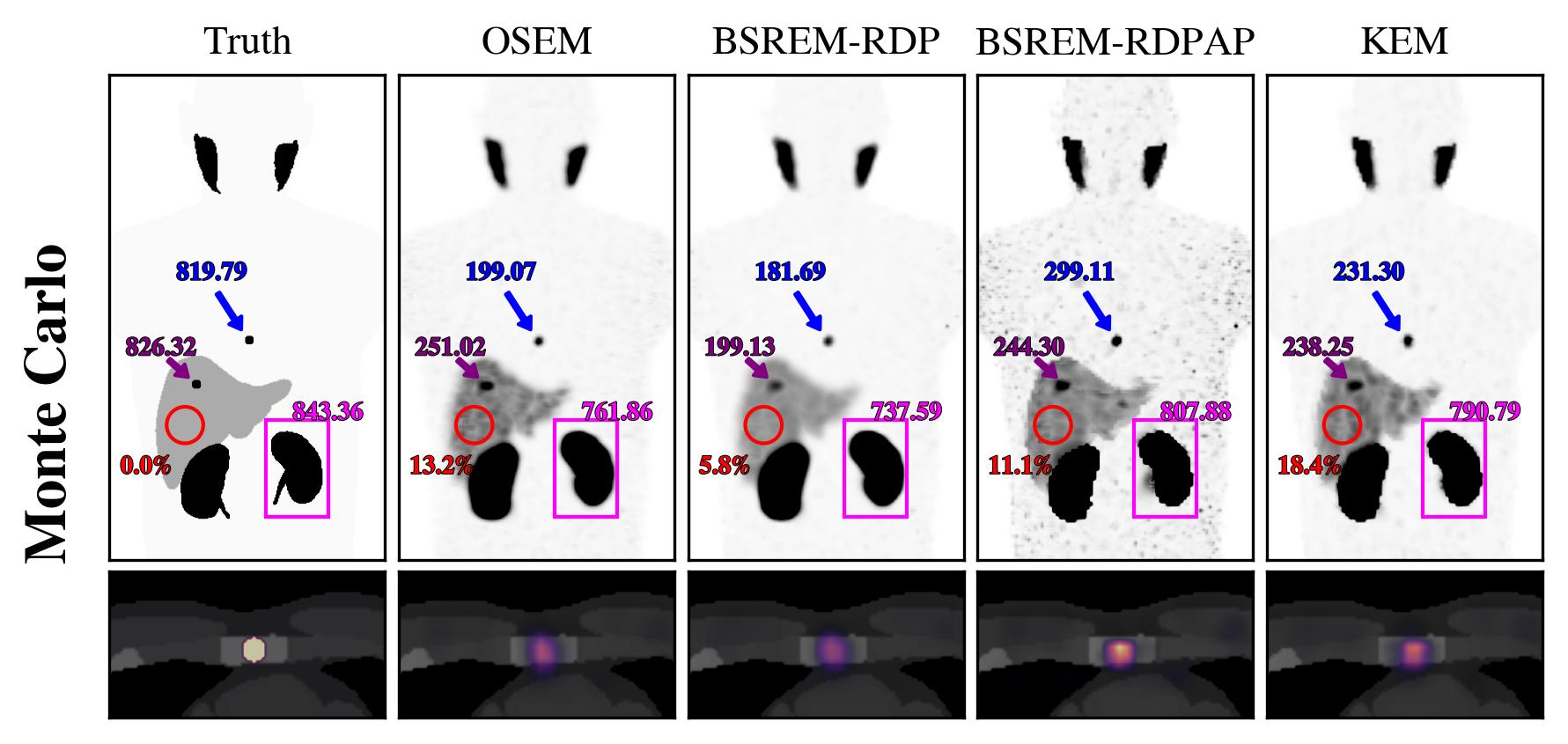}
   \caption{DICOM SPECT reconstruction of clinical data (top) and SIMIND SPECT reconstruction of Monte Carlo data (bottom). Greyscale images correspond to maximum intensity coronal projections of reconstructed images. The annotations correspond to: (i) the left kidney, contained inside a pink box, with the corresponding activity concentration in kBq/mL (ii) the location of the spherical mask in the liver used for noise quantification with the corresponding noise in percent, (iii) the location of the bone metastasis (blue arrow) with the corresponding activity concentration in kBq/mL, and (iv) the location of the liver lesion (purple arrow) inserted in the SIMIND data. Axial slices are shown below each greyscale image that correspond to the slice containing the bone metastasis; the SPECT activity is shown with a magma colormap superimposed on a Greyscale CT image.
   \label{fig:spect_recon_figure} 
    }  
    \end{center}
\end{figure}

For the clinical data, the left kidney was segmented using the CT-based TotalSegmentator model \cite{3D_segmenter}, and the bone metastasis in the sternum was segmented by first segmenting the lesion from the OSEM reconstruction, and then applying a threshhold mask greater than $0.15~$cm$^{-1}$ so that only voxels within the bone were considered. A spherical mask was used in the liver for both the simulated and clinical data for measurement of noise in each reconstruction algorithm; noise was quantified as a percent by computing the standard deviation of uptake divided by the mean of uptake in that region. Reconstructions of the clinical and simulated data are shown in Figure \ref{fig:spect_recon_figure}.\\

\subsection{PET}
PyTomography is capable of reconstructing real and simulated PET data in either sinogram mode or listmode. This example uses GATE Monte Carlo data and demonstrates the following capabilities of the library in PET reconstruction:
\begin{enumerate}
    \item LM-TOF reconstruction
    \item Random estimation via delayed coincidences and scatter estimation using the TOF single scatter simulation (TOF-SSS) technique
    \item Use of the DIPRecon algorithm to enhance reconstruction; the \verb|DIPRecon| class of PyTomography directly takes as input an instance of an \verb|nn.Module| of PyTorch, highlighting the library capabilities to interface directly with neural networks.
\end{enumerate}

 The simulated scanner consisted of $8$ rings of 338 mm radius with 32.25 mm ring spacing, and a TOF coincidence resolution of 550 ps; this fictional scanner is representative of a Siemens Biograph mMR with TOF capabilities. The digital phantom used for simulation in  GATE consisted of an ultra-high resolution $^{18}$F-FDG PET image, a T1-weighted MRI brain image, and a corresponding attenuation map \cite{brain_phantom}. The total uptake of 15.66~MBq was representative of a typical $^{18}$F-FDG distribution 1 hour post injection. The normalization $\eta$ was obtained via simulation of a thin cylindrical shell of radius $318~$mm that extended along the axial FOV of the scanner; normalization computation is available in PyTomography for basic polygonal PET geometries. In all cases, simulated data were reconstructed using a $204 \times 204 \times 154$ matrix size with voxel spacing of $1.25~$mm; 21 TOF bins were used with a total range of 30~cm. Before reconstruction, the ``randoms'' contribution to each listmode event was estimated from delayed coincidences; this process consisted of (i) binning the delayed coincidence events into a sinogram (ii) smoothing the sinogram to reduce noise, and (iii) converting back into LM form for use in LM reconstruction. The ``scatters'' contribution for each listmode event was estimated using the TOF single scatter simulation (SSS) capabilities of the library, based on Watson \cite{watson_Scatter}. The PET image used as a proxy for SSS was a reconstruction of the data via LM-TOF OSEM (40it, 1ss) without scatter estimation. Simulation and reconstruction were then performed in the following three ways:
\begin{enumerate}
    \item A 9 minute acquisition (high count) was simulated in GATE that included all relevant physical phenomenon in a typical PET scan. The data was then reconstructed using LM-TOF OSEM for up to 27 iterations with 3 subsets with attenuation, random, and scatter modeling. PSF modeling was implemented via 3 dimensional blurring in object space using a Gaussian kernel with a full width half max of 3~mm.
    \item The dataset above was subsampled into a 3 minute acquisition (low count), and an identical reconstruction was performed for up to 81 iterations with 1 subset.
    \item A low count reconstruction was performed without PSF modeling and used as an initial image for the DIPRecon algorithm; no PSF modeling was used in the initial image because (i) the DIP performs better on images with high frequency noise, and (ii) the DIPRecon algorithm already has noise reduction properties. The prior image $z$ used was the corresponding T1-weighted MRI image; the initial network prediction $\mathcal{N}(z|\theta_0)$ was obtained via fitting to the low count reconstruction. The neural network was a UNet \cite{UNet} with similar architecture to the original DIP paper \cite{deep_image_prior}; more information on the network structure and reconstruction hyperparameters can be found in the publicly available code. Although PSF modeling was not used to obtain the initial image, it was employed during the iterative process of the DIPRecon algorithm.
\end{enumerate} 

\begin{figure}[ht]
   \begin{center}
   \includegraphics[width=13cm]{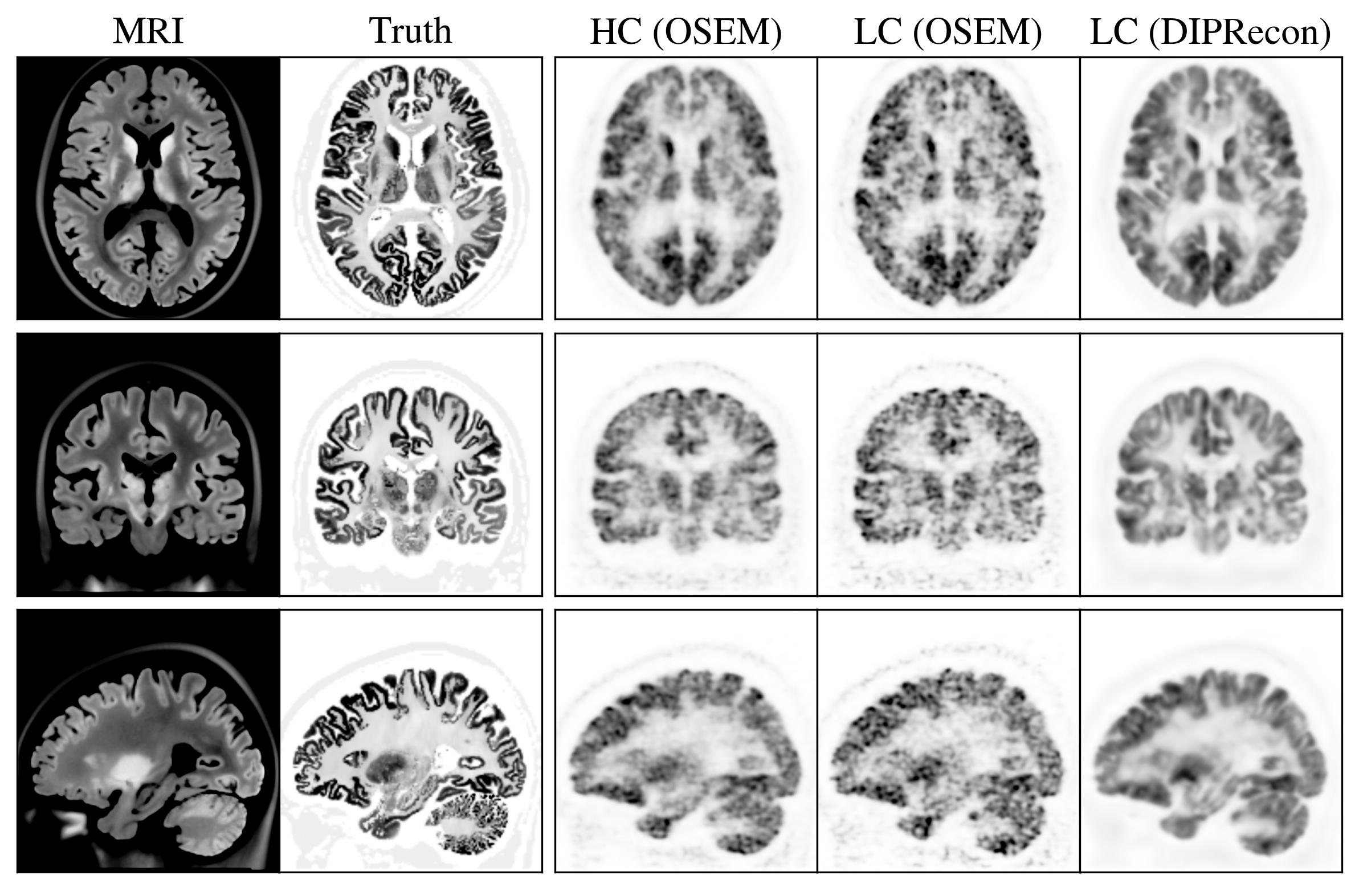} \includegraphics[width=13cm]{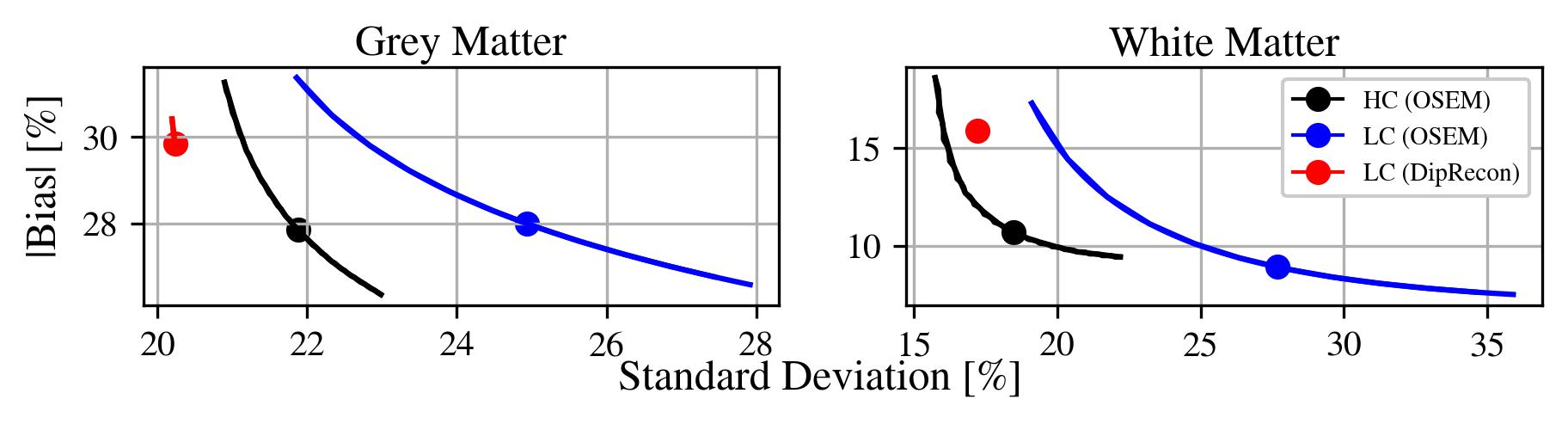}
   \caption{Top: PET reconstruction of GATE simulated data corresponding to a ultra-high resolution $^{18}$F-FDG PET/ T1-weighted MRI brain image \cite{brain_phantom}. From top to bottom: axial, coronal, and sagittal slices. From left to right: ground truth MRI (non contrast adjusted), ground truth PET uptake, high count OSEM reconstruction (18it/3ss shown), low count OSEM reconstruction (54it/1ss shown), and low count DIP reconstruction. Bottom: Variation in absolute bias and standard deviation throughout iterative reconstruction; points are plotted at the iteration corresponding to the displayed images. Subplots correspond to the grey matter region (left) and white matter region (right). The bias was initially positive and became negative for all three cases in white matter.
   \label{fig:pet_dip} 
    }  
    \end{center}
\end{figure}

Corresponding phantom reconstructions, as well as bias and noise in grey and white matter as a function of iteration number, are shown in Figure \ref{fig:pet_dip}. Required computation times are shown in Table \ref{tab:recon_times}.

\begin{table}[h]
    \centering
    \caption[asd]{Computational times for all use cases considered in this paper. Times required to open raw data are not considered. Times to construct SPECT system matrices and compute PET randoms in PyTomography are not considered since these are negligible. The notation ``s / it'' corresponds to the seconds per iteration for a given reconstruction algorithm, `SSS' refers to the single scatter simulation in PET, and `SM' refers to the creation of the system matrix. For the DIPRecon PET case, `IF' refers to the initial model fit required for the DIPRecon. All timing performed on the same GPU-optimized system except for the CASToR PET validation example, which was ran on a CPU-optimized system.\label{tab:recon_times}}
    \begin{tabular}{|c|c|}
        \hline
        \textbf{SPECT DICOM} & Computation Time\\
        \hline
        \hline
        OSEM (4it/6ss) & (1.32~s/it/bed) $\times$ 4it $\times$ 2 bed $=$ \textbf{10.6~s}\\
        BSREM-RDP (40it/6ss) & (1.13~s/it/bed) $\times$ 40it $\times$ 2 bed $=$ \textbf{90.4~s}\\
        BSREM-RDPAP (40it/6ss) & (1.21~s/it/bed) $\times$ 40it $\times$ 2 bed $=$ \textbf{96.8~s}\\
        KEM (10it/6ss) & (1.52~s/it/bed) $\times$ 10it $\times$ 2 bed $=$ \textbf{30.4~s}\\
        \hline
        \hline
        \textbf{SPECT SIMIND} & Computation Time \\
        \hline
        \hline
        OSEM (4it/6ss) & (1.07~s/it) $\times$ 4it $=$ \textbf{4.3~s}\\
        BSREM-RDP (40it/6ss) & (0.83~s/it) $\times$ 40it $=$ \textbf{33.2~s}\\
        BSREM-RDPAP (40it/6ss) & (0.86~s/it) $\times$ 40it $=$ \textbf{34.4~s}\\
        KEM (10it/6ss) & (1.07~s/it) $\times$ 10it $=$ \textbf{10.7~s}\\
        \hline
        \hline
        \textbf{PET GATE} & Computation Time\\
        \hline
        \hline
        High Count OSEM (27it/3ss) & $\begin{aligned} (21.1~\text{s/it})& \times 27\text{it} +30.9~\text{s (SM)} \\  + &925.9~\text{s (SSS)} = \textbf{1525.5~\text{s}} \end{aligned}$\\
        Low Count OSEM (80it/1ss) & $\begin{aligned} (6.69~\text{s/it})& \times 80\text{it} +30.8~\text{s (SM)} \\  + &1023.7~\text{s (SSS)} = \textbf{1589.8~\text{s}} \end{aligned}$\\
        Low Count DIPRecon (100it) & $\begin{aligned} (25.5~\text{s/it})& \times 100\text{it} +663.8~\text{s (IF)} \\  = &\textbf{3209.6~\text{s}} \end{aligned}$\\
        \hline
        \hline
        \textbf{SPECT Validation} & Computation Time\\
        \hline
        \hline
        PyTomo OSEM (2it/8ss) & (2.70~s/it) $\times$ 2it $=$ \textbf{5.4~s}\\
        STIR OSEM (2it/8ss) & \textbf{51542.2~s}\\
        \hline
        \hline
        \textbf{PET Validation} & Computation Time\\
        \hline
        \hline
        PyTomo OSEM (2it/21ss) & (59~s/it) $\times$ 2it $=$ \textbf{118~s}\\
        CASToR OSEM & \textbf{910~s}\\
        \hline
    \end{tabular}
\end{table}

\section{Impact}
In this work, the software architecture of PyTomography was demonstrated and use cases were explored on both SPECT and PET data. The efficient reconstruction times, shown in Table \ref{tab:recon_times}, can enable researchers to perform extensive studies on digital/physical phantom and patient data that may include multiple patients, noise realizations, and reconstruction algorithms. A detailed discussion on each use case is included to highlight example analyses that can be performed with the software.\\

A variety of algorithms were examined for $^{177}$Lu SPECT reconstruction; various quantitative metrics were estimated to highlight the differences between the available algorithms in the library. For both the simulated and clinical data, the algorithms yielding smallest to largest kidney concentration were BSREM-RDP, OSEM, KEM, and BSREM-RDPAP. The simulated data prediction from BSREM-RDPAP ($807.88~$kBq/mL) was closest to the ground truth ($843.36~$kBq/mL), suggesting that algorithms incorporating anatomical information may yield the most accurate kidney activity concentration. For both the simulated and clinical data, the order of algorithms yielding the lowest to highest noise were BSREM-RDP, BSREM-RDPAP, OSEM, and KEM; this is unsurprising as the RDP prior is designed to reduce noise. While BSREM-RDPAP has a lower noise than OSEM in each case, it also had higher uptake in the kidney: BSREM-RDPAP thus has simultaneous improvements in both bias and noise reduction. For the bone lesion, algorithms incorporating anatomical information (BSREM-RDPAP and KEM) were able to contain activity mainly inside the bone; this is demonstrated in the axial slices shown in Figure \ref{fig:spect_recon_figure}. These algorithms also yielded higher bone lesion activity concentration predictions than OSEM for both the simulated/clinical data, with the simulated data concentrations ($299.11~$kBq/mL for BSREM-RDPAP and $231.30~$kBq/mL for KEM) closer to the ground truth ($819.79~$kBq/mL) than OSEM ($199.07~$kBq/mL). For the liver lesion in the simulated data, however, use of BSREM-RDPAP actually yielded a smaller activity concentration prediction ($244.30~$kBq/mL) compared to OSEM ($251.02~$kBq/mL). This occured because there was no anatomical difference in the surrounding tissue in the liver metastasis, while there was for the bone metastasis. These results, which are consistent with previous studies \cite{hermes_spect_ap}, suggest that anatomical priors may yield more accurate activity concentrations for bone lesion dosimetry. While this example focused primarily on organ and lesion dosimetry, PyTomography users will be to assess the merits of various reconstruction algorithms on any task specific metric.\\

Example applications in $^{18}$F-FDG PET imaging demonstrated the capabilities of LM TOF SSS modeling and LM-TOF reconstruction using the OSEM and DIPRecon algorithms. As apparent in Figure \ref{fig:pet_dip}, reconstruction of the low count data using DIPRecon was able to reduce noise in the grey matter regions while preserving cortical structures. While use of the DIPRecon on the low count data resulted in a slight increase in bias in the grey matter region, the corresponding noise bias curve demonstrates less noise than the high count reconstruction using OSEM. Extensive research on the validity of the DIPRecon method may seek to explore, for example, the effects of different neural network architectures, hyperparameters, and misalignment between the PET/MRI images. Since PyTomography directly interfaces with PyTorch, the process of designing and testing neural networks and optimization procedures for the DIPRecon method is straightforward. Users who wish to test different network/optimization configurations are encouraged to view the DIPRecon tutorial available on the documentation website.\\

While this paper has demonstrated implementations for parallel collimator SPECT and LM-TOF PET, there exist many other modalities not presently included in the software, including (but not limited to) diverging/converging/pinhole SPECT, various forms of computed tomography, (CT), magnetic resonance (MR) imaging, and compton camera (CC) imaging. It is hoped that the modular structure and flexibility of PyTomography will encourage contributions from various research groups interested in particular imaging systems or reconstruction algorithms. Since the library is open source, newly developed functionality can be easily tested and verified by many independent research groups with their own private data. The immediate goals for future development in PyTomography at the time of publication include:
\begin{enumerate}
    \item Expanded support for clinical PET data.
    \item Support for Monte Carlo scatter correction in SPECT reconstruction.
    \item Support for CT reconstruction
    \item Development of a 3D Slicer \cite{3DSlice} extension.
\end{enumerate}

\section{Conclusions}
This work describes the python library PyTomography and highlights specific use cases in SPECT and PET imaging. The software architecture facilitates the development of different imaging modalities and likelihoods that can all interface with the same reconstruction algorithms. The goal of this research was to create a simple-to-use and computationally-efficient library for medical image reconstruction, where system modeling and reconstruction techniques are implemented, shared, and explored by experts in the community.

\section*{Acknowledgements}
\label{}
This work was supported by the Natural Sciences and Engineering Research Council of Canada (NSERC) CGS D Award 569711, NSERC Discovery Grants RGPIN-2019-06467 and RGPIN-2021-02965, as well as computational resources and services provided by Microsoft AI for Health.

\appendix

\section{Validation of PyTomography}

This section contains three examples used to validate PyTomography against other reconstruction software. Reconstruction times for cases 1 and 3 are shown in Table \ref{tab:recon_times}.

\begin{figure}[ht]
   \begin{center}
   \includegraphics[width=13cm]{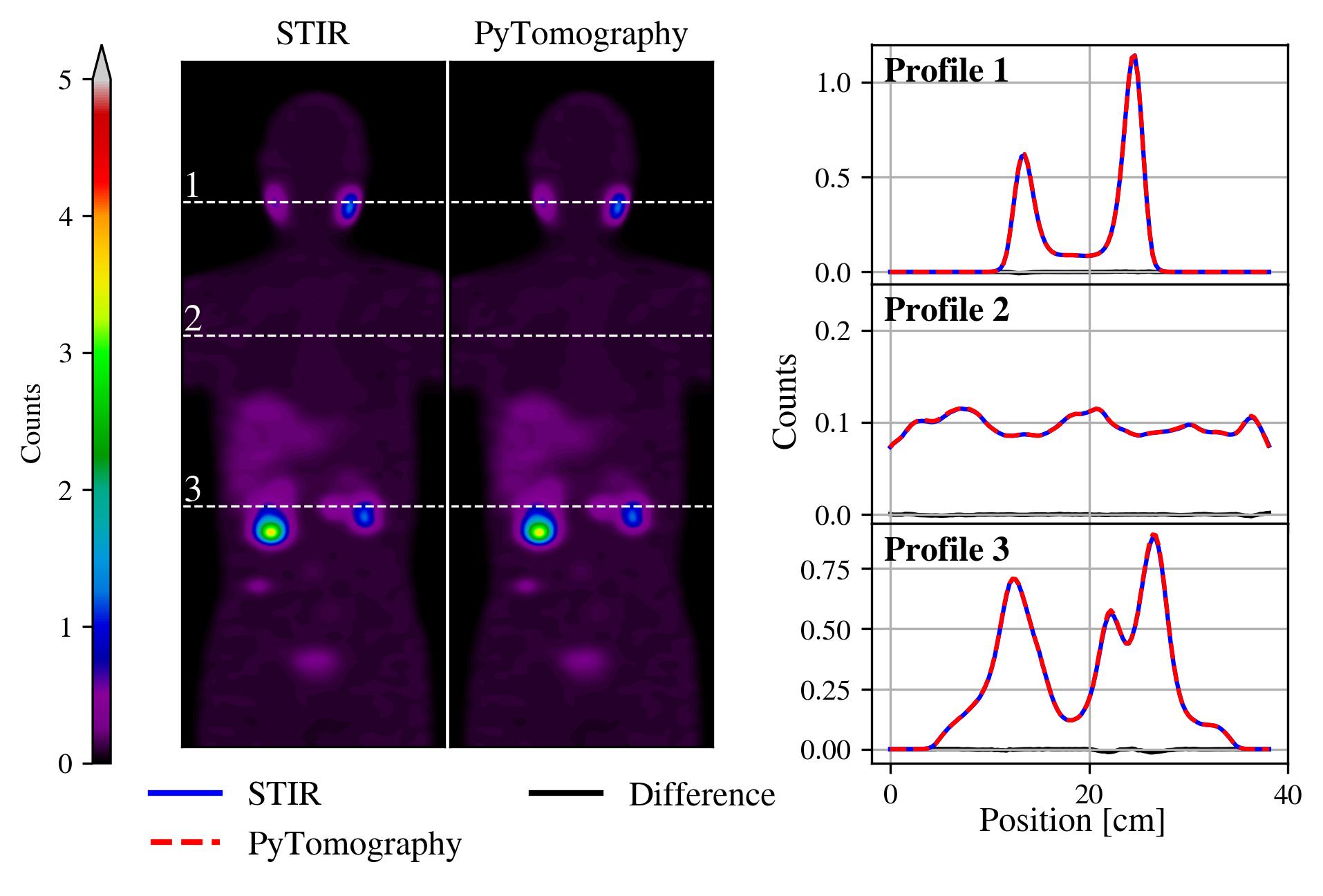}
   \caption{Central coronal slices (left) and 1D profiles (right) of reconstructed SIMIND SPECT projection data using OSEM (2it/8ss) with attenuation/PSF modeling; images were reconstructed using STIR and PyTomography. 1D profiles correspond to the horizontal lines shown on the coronal slices. 
   \label{fig:py_vs_stir} 
    }  
    \end{center}
\end{figure}

\begin{figure}[ht]
   \begin{center}
   \includegraphics[width=13cm]{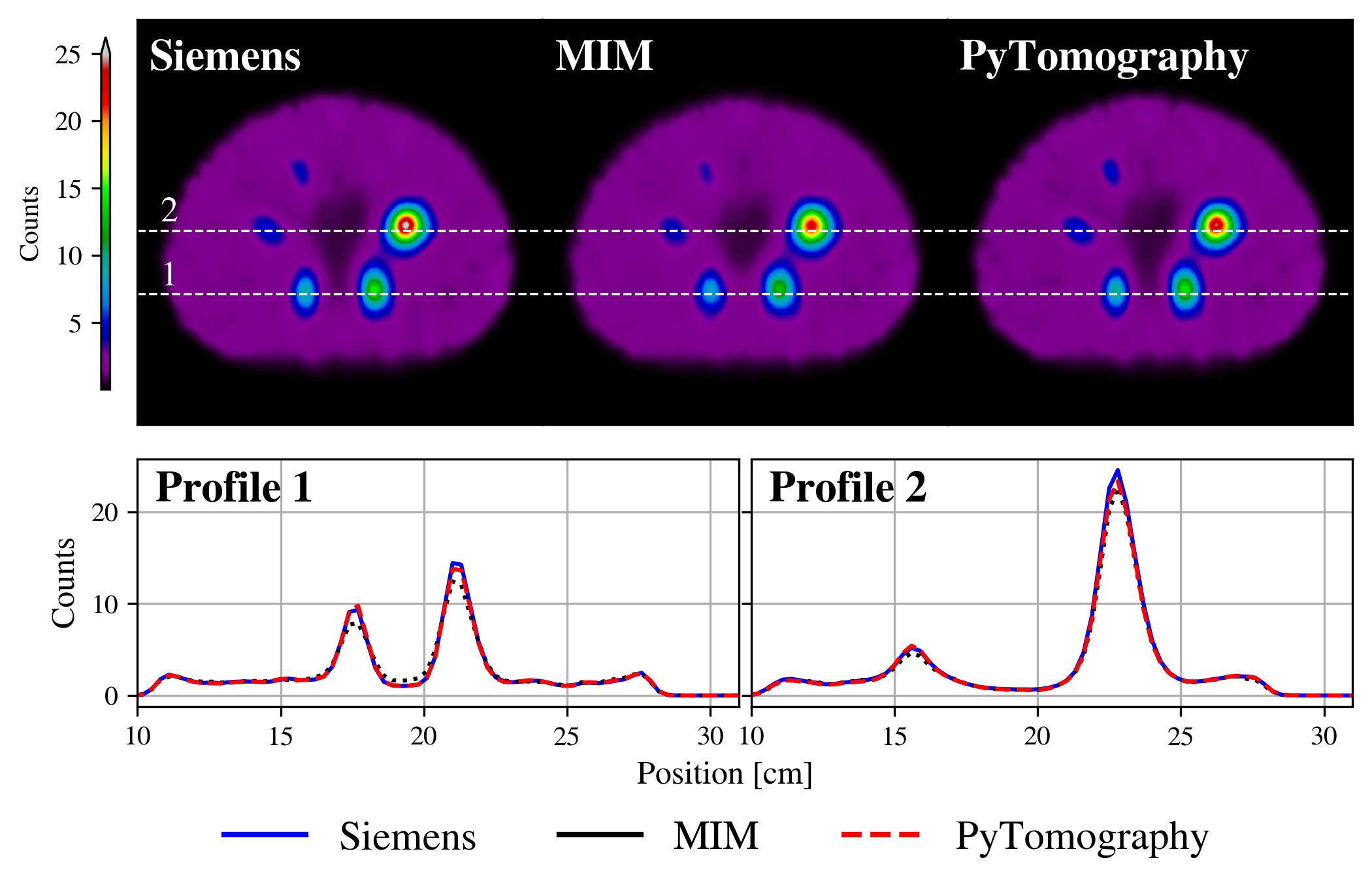}
   \caption{Sample axial slices (top) and 1D profiles (bottom) of a $^{177}$Lu NEMA phantom reconstructed using OSEM (4it/8ss) with attenuation/PSF modeling and triple energy scatter correction. Reconstructions were performed using (i) the Siemens scanner software, (ii) MIM SPECTRA Reconstruction, and (iii) PyTomography. 1D profiles correspond to the horizontal lines shown on the axial slices. 
   \label{fig:dicom} 
    }  
    \end{center}
\end{figure}

\begin{figure}[ht]
   \begin{center}
   \includegraphics[width=13cm]{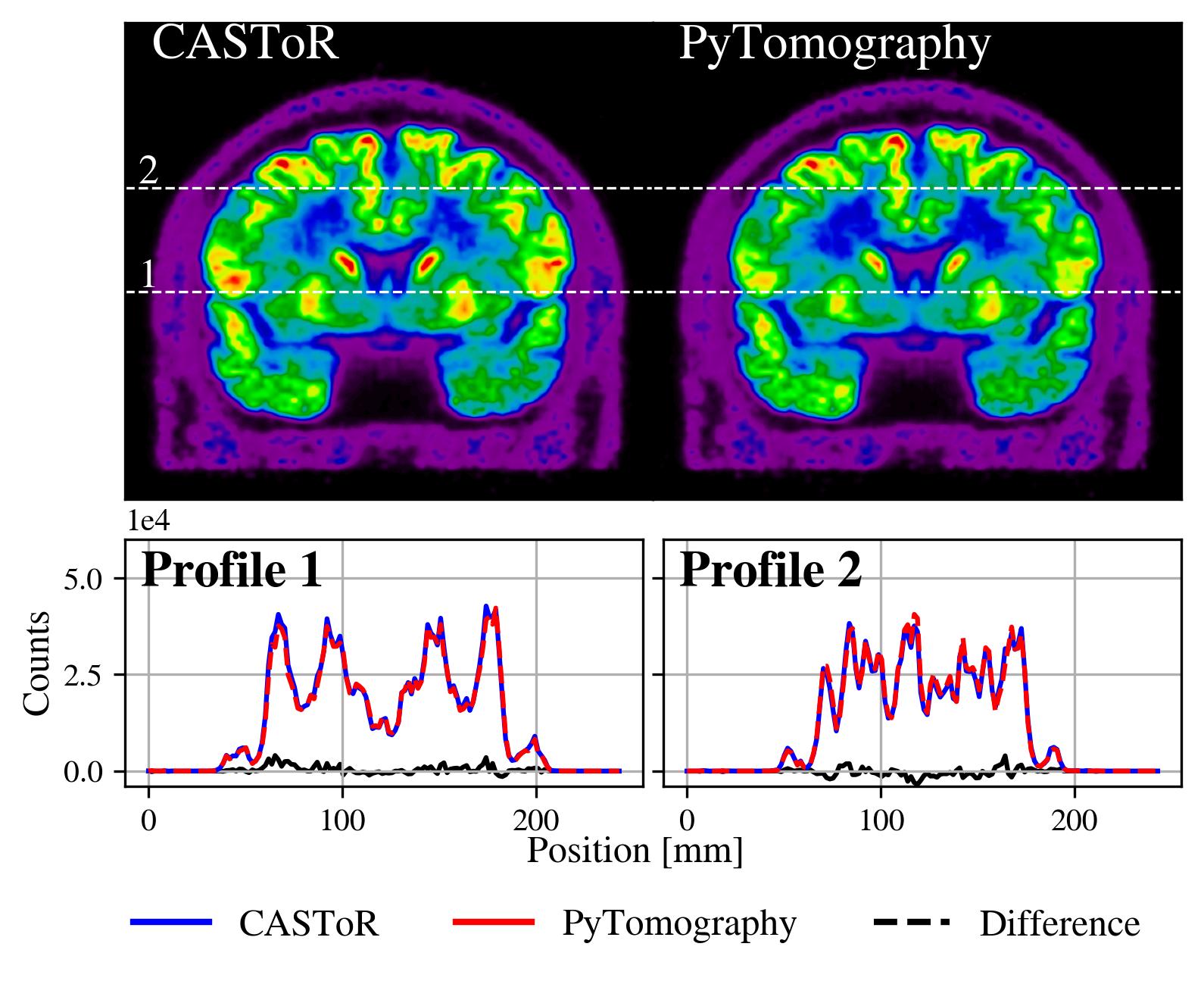}
   \caption{Sample coronal slices (top) and 1D profiles (bottom) of the high resolution brain phantom reconstructed using CASToR and PyTomography with LM TOF OSEM (2it/21ss). The two images were normalized such that the sum of their voxels was identical. 1D profiles correspond to the horizontal lines shown on the coronal slices. The percent difference in grey and white matter mean uptake between the two software packages were 0.90\% and 1.14\% respectively.
   \label{fig:pet_validation} 
    }  
    \end{center}
\end{figure}

\begin{enumerate}
    \item \textbf{Monte Carlo SPECT Validation With STIR}: Radioactivity concentrations in a digital phantom were selected to correspond to a 24-hour post injection prostate cancer patient  (1700 MBq total activity) imaged via the prostate-specific membrane antigen (PSMA)-targeting radiopharmaceutical, $^{177}$Lu-PSMA-617. SPECT acquisition was simulated in SIMIND with dimensions $128 \times 384$, a pixel size of $0.48~\text{cm} \times 0.48~\text{cm}$, 120 projections, an acquisition time of 15~s/projection, and use of Siemens medium energy collimators. Data were reconstructed with OSEM (2 iterations, 8 subsets) using (i) STIR and (ii) PyTomography. Coronal slices and 1D profiles of reconstructions are shown in Figure \ref{fig:py_vs_stir} and are nearly identical. The image noise, quantified by the standard deviation of predicted counts in the liver, was 0.789 counts (STIR) and 0.793 counts (PyTomography). The mean predicted counts in the kidneys were 1.764 (STIR) and 1.761 (PyTomography).
    \item \textbf{Clinical SPECT Validation With Siemens and MIM}: A NEMA phantom was filled with $^{177}$Lu at a 10:1 source to background ratio and scanned at 96 projections ($128 \times 128$ with a resolution of $4.8\text{mm} \times 4.8\text{mm}$) for 15~s per projection using a Siemens Symbia T Series SPECT/CT scanner with Medium Energy collimators. Reconstruction was performed with OSEM (4 iterations, 8 subsets) with attenuation/PSF modeling and TEW scatter correction, using (i) the manufacturers software: Siemens Symbia Intevo Excel v9.0.31.3, (ii) MIM SPECTRA v7.2.1, and (iii) PyTomography. Reconstructions and corresponding 1 dimensional profiles are shown in Figure \ref{fig:dicom}; the PyTomography and Siemens reconstructions are qualitatively more similar than the MIM reconstruction. 3D slicer was used to obtain sample segmentations of (i) the largest NEMA sphere, (ii) the central cold portion, and (iii) the warm background region. The mean counts in the largest NEMA sphere were 14.07 (Siemens), 12.96 (MIM), and 13.49 (PyTomography). The mean counts in the central cold portion were 0.732 (Siemens), 0.704 (MIM), and 0.688 (PyTomography). The image noise, quantified by the standard deviation of predicted counts in the warm region, were 0.196 (Siemens), 0.168 (MIM), and 0.179 (PyTomography). 
    \item \textbf{Monte Carlo PET Validation With CASToR}: The ultra-high resolution $^{18}$F-FDG PET/ T1-weighted MRI brain image \cite{brain_phantom} used in the PET example was simulated in GATE without random, scatter, attenuation, or positron range modeling to obtain a ultra high count dataset consisting only of primary events. The data were reconstructed using LM TOF OSEM (2 iterations, 21 subsets) with both PyTomography and CASToR. Resulting reconstructions are shown in Figure \ref{fig:pet_validation}. The mean activity ratio in grey to white matter was 2.05 (PyTomography) and 2.03 (CASToR). The noise in the image was estimated as the standard deviation of counts in white matter divided by the mean activity in white matter; the values were 0.329 (PyTomography) and 0.321 (CASToR). CASToR reconstruction was ran on a separate HP Z8 Computer Desktop with 2 Intel Xeon Gold 6136 CPUs at 3.00 GHz with a total of 48 CPU cores; this separate system was selected because it permitted use of more CPU cores in parallel during reconstruction, which is what CASToR is optimized for.
\end{enumerate}

The images shown in this section are displayed using the ``nipy\_spectral'' colormap of matplotlib because it has significant contrast. Since this colormap is neither perceptually uniform nor monotonically increasing in lightness, images in this section should be viewed in color. 

\clearpage








\end{document}